\documentclass[lettersize,journal]{IEEEtran}

\usepackage{amsmath,amsfonts}
\usepackage{algorithmic}
\usepackage{algorithm}
\usepackage{array}
\usepackage[caption=false,font=normalsize,labelfont=sf,textfont=sf]{subfig}
\usepackage{textcomp}
\usepackage{stfloats}
\usepackage{url}
\usepackage{verbatim}
\usepackage{graphicx}
\usepackage{cite}
\usepackage{bm}
\usepackage{textgreek}
\usepackage{amssymb}
\usepackage{cuted}
\usepackage{color}

\allowdisplaybreaks[4]
\begin{document}

\title{Joint Semantic Transmission and Resource Allocation for Intelligent Computation Task Offloading in MEC Systems}

\author{Yuanpeng Zheng,~\IEEEmembership{Student Member,~IEEE,} Tiankui Zhang, ~\IEEEmembership{Senior Member,~IEEE,} Xidong Mu, ~\IEEEmembership{Member,~IEEE,} Yuanwei Liu, ~\IEEEmembership{Fellow,~IEEE,} and Rong Huang
\thanks{This work is supported by the National Natural Science Foundation
of China (NSFC) under Grant 62371068. 
(Corresponding author: Tiankui Zhang)}
\thanks{Yuanpeng Zheng, Tiankui Zhang are with the School of Information and Communication Engineering, Beijing University of Posts and Telecommunications, Beijing 100876, China (e-mail: \{zhengyuanpeng, zhangtiankui\}@bupt.edu.cn).

Xidong Mu is with with Queen's University Belfast, Belfast, BT3 9DT, U.K. (e-mail: x.mu@qub.ac.uk).

Yuanwei Liu is with the Department of Electricaland Electronic Engineering, the University of Hong Kong, Hong Kong (e-mail: yuanwei@hku.hk). 

Rong Huang is with the China Unicom Research Institute, Beijing, China. (email: huangr27@chinaunicom.cn).
}
}

\maketitle

\begin{abstract}
Mobile edge computing (MEC) enables the provision of high-reliability and low-latency applications by offering computation and storage resources in close proximity to end-users. 
Different from traditional computation task offloading in MEC systems, the large data volume and complex task computation of artificial intelligence involved intelligent computation task offloading have increased greatly.
To address this challenge, we propose a MEC system for multiple base stations and multiple terminals, which exploits semantic transmission and {early exit of inference}.
Based on this, we investigate a joint semantic transmission and resource allocation problem for maximizing system reward combined with analysis of semantic transmission and intelligent computation process.
To solve the formulated problem, we decompose it into communication resource allocation subproblem, semantic transmission subproblem, and computation capacity allocation subproblem. 
Then, we use 3D matching and convex optimization method to solve subproblems based on the block coordinate descent (BCD) framework. 
The optimized feasible solutions are derived from an efficient BCD based joint semantic transmission and resource allocation algorithm in MEC systems. 
Our simulation demonstrates that: 1) The proposed algorithm significantly improves the delay performance for MEC systems compared with benchmarks;
2) The design of transmission mode and early exit of inference greatly increases system reward during offloading;
and 3) Our proposed system achieves efficient utilization of resources from the perspective of system reward in the intelligent scenario.
\end{abstract}


\section{Introduction}
Mobile edge computing (MEC) {can reduce the load on mobile networks, decrease transmission delays, and meet service quality requirements by providing communication, computation, and storage services to nearby mobile devices. 
Concurrently, the industry has seen the rise of numerous artificial intelligence applications, supported by high-speed, low-latency mobile cellular networks as infrastructure. 
These applications encompass a range of intelligent recognition tasks within the Internet of Things (IoT), the Internet of Vehicles (IoV), and other related areas.}
Consequently, the MEC system{s} are required to accommodate a large number of intelligent computation tasks from mobile terminals.
Different from traditional computation task offloading, intelligent computation task offloading introduce artificial intelligence inference models which have complex computation logic and demand significant resources.
On the one hand, intelligent models in task offloading \cite{ref2} brings nonlinear computation processes \cite{ref3} {which represents a type of computation process where the relationship between the required computation cost and the input data size is nonlinear.} 
On the other hand, {the more complex} data characteristics in various scenarios will cause uplink data volume of tasks transmitted by terminals to be large \cite{ref31}.
These factors both bring new challenges to implementation of MEC in the intelligent scenarios.

Introducing intelligent computation task offloading faces a series of problems owing to the constrained computation resources of the MEC system.
Various types of training and inference tasks of neural networks are included such as deep neural network (DNN), convolutional neural network and other intelligent classification recognition related tasks.
The structure of the computation model for such tasks is very complex which often requires more computation capacity. 
Therefore, resource pressure on MEC is greater \cite{ref32} and it is more difficult to analyze computation capacity requirements while allocating resources \cite{ref10}.

{Moreover, intelligent computation tasks typically involve more complex types of data, such as images, videos and multi-modal data \cite{ref33}, resulting in larger data volume compared to traditional situations.
The larger amount of offloaded data brings greater data transmission pressure to the computation offloading process of the MEC systems.
This type of transmission tasks is difficult to meet system performance requirements using traditional methods.}
Therefore, it is important to focus on the communication traffic load and delay performance caused by changes in the data volume of the task while considering resource allocation. 

\subsection{Prior Works}
At this stage, the relevant work pertaining to computation offloading and resource allocation of MEC primarily commences from various traditional computation scenarios and encompasses the consideration of efficient algorithms \cite{ref4,ref5,ref6,ref7,ref8,ref9}.
Among them, Shi \textit{et al.}\cite{ref4} considered the joint optimization of task offloading and resource allocation to efficiently fulfill service requests in the MEC network with spatial-temporal dynamics.
Some work \cite{ref5,ref6} focused on IoV and IoT scenarios and presented an efficient {task offloading} and resource allocation scheme in MEC systems.
Feng \textit{et al.}\cite{ref7} considered the scenario of federated learning and proposed a {computation offloading} and resource allocation architecture based on the heterogeneous mobile framework. 
In \cite{ref8,ref9}, authors {proposed energy consumption and resource allocation optimization schemes for task offloading of special unmanned aerial vehicle (UAV) scenarios in MEC systems.}
A number of research about the field of intelligent computation and {lightweight models} has emerged in recent years. 
Some of the work currently analyzed models in intelligent computation tasks involving communication and {decentralized computation method}\cite{ref1,ref3,ref11,ref12,ref35,ref13,ref18}.
Some research \cite{ref11,ref3} analyze the structure of neural networks and propose different architectures to improve computational efficiency {to make them more lightweight}.
{To reduce communication overhead and computation overhead in IoT systems, Ayad \textit{et al.}\cite{ref12} introduced a modified split learning system that includes an autoencoder and an adaptive threshold mechanism. 
Yoon \textit{et al.}\cite{ref1} developed a lightweight natural image matting network with a similarity-preserving knowledge distillation which is effective for mobile applications. 
Bai \textit{et al.}\cite{ref35} proposed a novel algorithm called neural ensemble to solve the DNN ensemble formation problem  considering the device heterogeneity, computing resource limitation, and service deadline of edge computing systems.}
Kang \textit{et al.}\cite{ref13} proposed an aerial image transmission paradigm for scene classification tasks considering lightweight model deployment of UAV edge systems. 
{Ren \textit{et al.} \cite{ref18} proposed a new semantic communication network to extract semantic information from images which provides efficient and high-performance image transmission for IoT device.}

As a step forward, there are ongoing research exploring the implementation of intelligent computation task offloading in MEC systems.
Some work mainly focus on the deployment of intelligent models at edge and related computation offloading and resource allocation research \cite{ref2,ref15,ref16,ref17,ref14,ref19,ref20,ref21,ref22,ref23}.
Zheng \textit{et al.}\cite{ref15} considered semantic extraction tasks and performed dynamic multi-time scale resource optimization in MEC systems. 
In detail, Teerapittayanon \textit{et al.}\cite{ref2} combined edge computing and the deployment of neural networks incorporating early exit network to allow fast and local inference. 
Simultaneously, authors in \cite{ref16,ref17} proposed a DNN inference framework {of cloud-edge-device synergy in MEC systems to improve performance of mobile intelligent services.} 
{By combining edge and intelligent computation, Dong \textit{et al.}\cite{ref14} introduced an offloading framework based on a large language model for MEC.}
In \cite{ref20,ref23}, the work focused on the deployment of task-oriented communication scheme on MEC and propose task offloading and resource allocation methods considering intelligent requirements. 
Fan \textit{et al.}\cite{ref19,ref21,ref22} considered edge-assisted machine learning task inference, model deployment and model segmentation in MEC systems and proposed different resource management schemes.

\subsection{Motivation and Contribution}
As summarized above, prior research have revealed some novel scenarios in MEC systems, and studied the nonlinearity of intelligent models and the deployment in mobile networks.
Based on these research, some work has considered the increasingly complex intelligent computation task offloading scenarios in MEC systems, including DNN deployment, intelligent requirements and task inference mechanisms.
{These increasingly complex intelligent scenarios will bring varying degrees of resource problems to MEC systems due to the complexity of their network deployment and the diversity of computation forms.}
However, the resource pressure caused by intelligent computation process and data characteristics have not been resolved.
Aiming at the two problems in above scenarios, i.e., complex task computation and large transmission data volume, this paper makes a pioneering work to optimize the influence on resource allocation based on specific tasks during offloading.
Specifically, {a common early exit of inference (EEoI) mechanism \cite{ref3} is introduced into the MEC systems which is a type of modified neural network performing early exit to output inference results while meeting accuracy requirements.
The EEoI simplifies the nonlinear computation process of neural networks and can greatly reduce resource consumption for complex task computation.}
Meanwhile, semantic transmission mechanism, where terminals can perform semantic extraction and compression considering the heterogeneity of data, is promising for reducing large transmission data volume pressure motivated by \cite{ref34}.
{To the best knowledge of the authors, the resource allocation problem of introducing the above two mechanisms in MEC systems has not been studied.}
Against the above background, we construct a MEC system based on above two designs and attempt to provide a optimization algorithm to achieve efficient resource allocation for intelligent computation task offloading.
We make the following contributions in this paper:
\begin{itemize}
  \item{
    We propose a MEC system for multiple base stations (BSs) and multiple terminals, which exploits semantic transmission and {EEoI}.
    Taking the image task as an example, we model the semantic transmission process and theoretically analyze the nonlinear computation overhead caused by EEoI, which improve offloading efficiency but lead to heterogeneity in inference computation.
    Furthermore, we propose a semantic transmission and resource allocation optimization problem for maximizing the system {reward} based on delay.
  }
  \item{
    We iteratively solve the formulated mixed nonlinear integer programming problem through block coordinate descent (BCD) based algorithm.
    Specifically, we decompose the problem into communication resource allocation subproblem, semantic transmission subproblem and computation capacity allocation subproblem, where the first subproblem is solved by 3D matching and the remaining subproblems are solved through convex optimization.
    Then we propose a BCD based joint semantic transmission and resource allocation algorithm in MEC systems to achieve the maximization of system reward.
  }
  \item{
    We conduct a simulation to verify that the proposed architecture is suitable for intelligent computation task offloading in MEC systems.
    Numerical results show that the proposed algorithm significantly improves the delay performance apparently compared with benchmarks.
    We also find that the design of transmission mode and EEoI greatly increases system reward during offloading.
    Moreover, the proposed system achieves efficient utilization of resources from the perspective of system reward in the intelligent scenario.
  }
\end{itemize}

\subsection{Organization and Notation}
The rest of this paper is organized as follows. Section II presents the framework of MEC system for multiple BSs and multiple terminals, which exploits semantic transmission and {EEoI}.
In Section III, the reward maximization problem is formulated, and a BCD-based iterative algorithm is proposed to solve the resulting non-convex problem. 
The performance of the proposed algorithm is evaluated by the simulation in Section IV, which is followed by the conclusions in Section V.
The main notations are shown as TABLE I.

\section{System Model}
\begin{figure*}[!t]
  \centering
  \includegraphics[scale = 0.6]{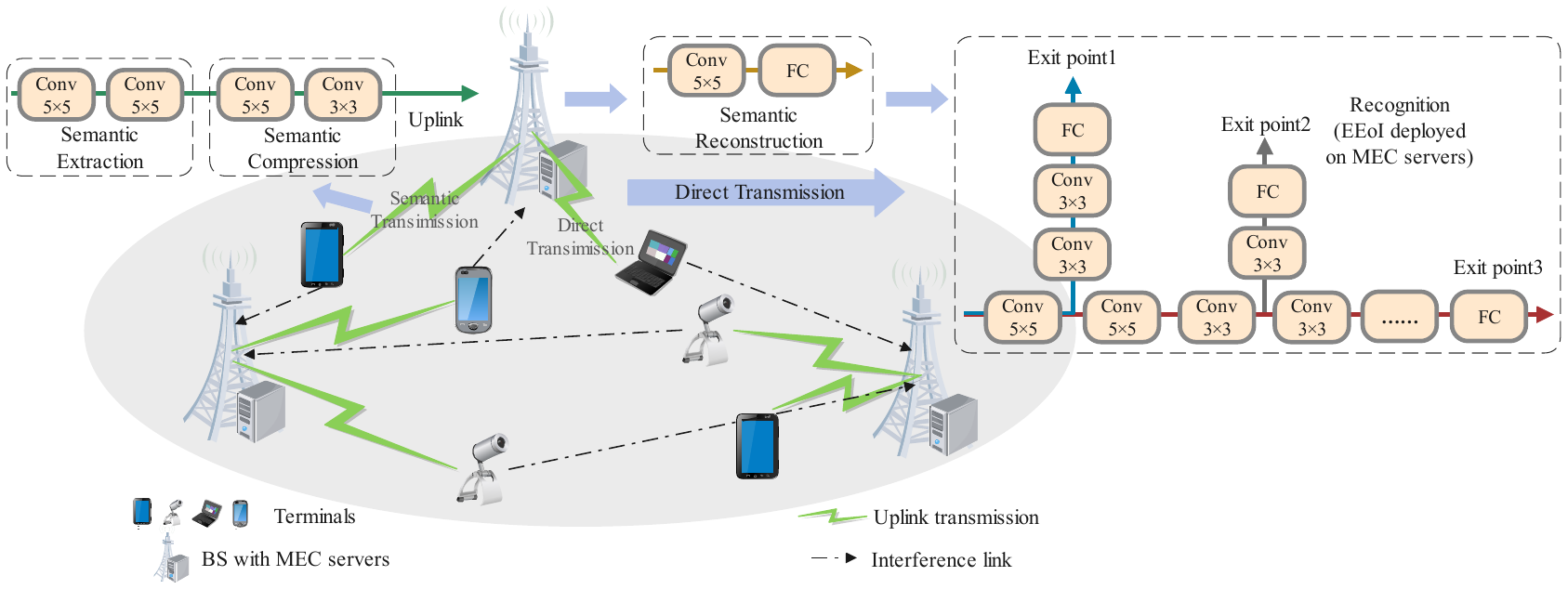}
  \caption{The system scenario of image semantic transmission.}
  \label{fig_1}
\end{figure*}

We consider the image semantic transmission scenario of mobile cellular network in the industrial Internet as shown in Fig. \ref{fig_1}. 
MEC servers are implemented on small base stations (SBSs) to establish the MEC systems, which is represented as ${\mathcal{K}} = \{1,...,k,...K\}$. 
There are image acquisition terminals in network coverage area, which is denoted as ${\mathcal{U}} = \{1,...,u,...,U\}$. 
{Since terminals have limited computational power, they cannot meet delay requirements when processing complex recognition tasks.}
Intelligent computation tasks they generate, i.e., ${\mathcal{I}} = \{1,...,i,...,I\}$ need to be offloaded to MEC servers for inference computation. 
{We assume that an unified architecture of AI model is adopted in our system, i.e., only one type of AI model can be requested by terminals at a time, and each task can always be processed with the corresponding deployed intelligent service on MEC servers.}
However, huge transmission and computation pressure of SBS and MEC servers caused by $\mathcal{I}$ need to be considered. 
It is necessary to deploy semantic transmission considering compression and extraction, and the EEoI mechanism that adds the early exit points, {which are trained early exit thresholds of inference divided according to task type to reduce computational overhead}, into neural networks according to data quality and accuracy requirements\cite{ref3}.
It will also lead to heterogeneity in MEC inference computation and nonlinearity in the computation model which needs to be considered.

\begin{table}[h]
  \renewcommand\arraystretch{1.5}
  \caption{Main Symbol and Variable List}
  \centering
  \fontsize{8}{7}\selectfont
  \begin{tabular}{|p{1.0cm}|p{6.25cm}|}
  \hline
  {\textbf{Notation}} & \textbf{Description} \\
  \hline
  $B$ & Bandwidth of a subcarrier  \\
  \hline
  $F_k$ & Computing capacity of each MEC  \\
  \hline
  $f^L_u$ & Computing capacity of each terminal  \\
  \hline
  $\tau_i$ & Delay limit for task $i$\\
  \hline
  $x_{uk}$ & Indicator of whether terminal $u$ is associated with MEC system $k$ \\
  \hline
  $z_{ui}$ & Indicator of whether task $i$ generated by terminal $u$ \\
  \hline 
  $\rho^n_{uk}$ & Indicator of whether subcarrier $n$ allocated to user $u$ associated with MEC system $k$\\
  \hline
  $r_{u}$ & Uplink transmission rate of terminal $u$\\
  \hline 
  $f_{u}$ & The computing capacity allocated to terminal $u$ by the MEC system\\
  \hline
  $t_{u}^{comm}$ & Transmission delay of terminal $u$ in the wireless link\\
  \hline
  $t_{u}^{comp}$ & Computation delay incurred by terminal $u$\\
  \hline
  $s_i$ & Data size of task $i$\\
  \hline
  $c_i$ & Computation amount required for recognition of task $i$\\
  \hline
  $\varepsilon_i$ & Compression ratio of task $i$\\
  \hline
  $F_R(\cdot)$ & Function of data size, exit point and computation amount required of recognition process\\
  \hline
  $F_E(\cdot)$ & Function of data size and computation amount required of semantic extraction and compression process\\
  \hline
  $F_C(\cdot)$ & Function of data size and computation amount required of semantic reconstruction process\\
  \hline
  $C_1,C_2$ & Parameters to adjust the value of system reward\\
  \hline
  $R_{ui}$ & The reward based on weighted delay of terminal $u$ with task $i$\\
  \hline
\end{tabular}
\end{table}

In our model, each terminal $u$ can be associated with the MEC system $k$ through the wireless channel provided by the BS. 
When task $i$ is generated, the terminal first decides whether to perform semantic extraction and compression based on the task information. 
Then, it offloads the task based on association and channel allocation.
After that, task $i$ is decided whether to perform semantic reconstruction according to the terminal processing situation by MEC servers.
{We assume that the task accuracy after EEoI has been trained to meet the system requirements considering the exit point $m_i$, semantic reconstruction and channel distortion.}
The semantic extraction, compression and reconstruction processes are deployed through an end-to-end convolutional neural network (CNN) and EEoI mechanism is set in recognition process as shown in Fig. \ref{fig_1}.

\subsection{Intelligent Computation Task Model}
We consider each terminal $u$ has a task $i$ and denote $z_{ui} \in \{0,1\}$ as the indicator variable of the task $i$ of user $u$. 
Specially $z_{ui} = 1$ when the task of user $u$ is $i$, and obviously $\sum_{i\in \mathcal{I}}z_{ui} \leq 1,\forall u \in \mathcal{U}$. 
The task $i$ has the following attributes: 
1) $s_i$, the task data size; 
2) $c_i$, the computation amount required for recognition; 
3) $\varepsilon_i$, compression ratio of the task transmitting via semantic extraction and compression which satisfies $\varepsilon_i \geq 1,\forall i\in \mathcal{I}$; 
4) $\tau_i$, the maximum delay that can be tolerated in task processing; 
5) $M_i$, the task priority which represents the importance of the task.

Note that the task data size $s_i$ satisfies $s_i \in [S^{LB},S^{UB}],\forall i, \label{eq2-1}$ 
where $S^{LB}$ and $S^{UB}$ are the lower bound and upper bound of the data size (e.g. the image size captured by terminal using low ang high solution modes), respectively.

\subsection{Task Transmission Model}
In our system, each BS is associated with the terminals through orthogonal frequency-division multiple access (OFDMA). We assume that the set of available subcarriers for each BS is ${\mathcal{N}} = \{1,...,n,...,N\}$ and the bandwidth of each subcarrier is $B$. 
Due to the orthogonality of the channel, it can be concluded that there is no interference within each BS but there is co-channel interference between BSs.
We define the offloading indicator vector as $\bm{x} = \{x_{uk}\in\{0,1\}\}_{u \in \mathcal{U}, k \in \mathcal{K}}$, where $x_{uk} = 1$ when terminal $u$ is associated with MEC system $k$. 
The subcarrier allocation indicator vector is denoted as $\bm{\rho} = \{\rho^n_{uk}\in \{0,1\}\}_{u \in \mathcal{U}, k \in \mathcal{K}, n\in \mathcal{N}}$, where $\rho^n_{uk} = 1$ when subcarrier $n$ is allocated to terminal $u$ associated with the MEC system $k$.
In that way, the signal to interference plus noise ratio (SINR) is given by 
\begin{equation}
  \label{eq2-2}
  \it{\Phi}^n_{uk} = \frac{P^n_{uk} g^n_{uk}}{\sum\limits_{c\in \mathcal{K},c\neq k}\sum\limits^{U}_{u' = 1}\rho^n_{u'c}g^n_{u'c}P^n_{u'c} + BN_0 },\forall u,k,n, 
\end{equation}
where $P^n_{uk}$ denotes transmit power from terminal $u$ to the MEC system $k$, $g^n_{uk}$ represents wireless channel gain between terminal $u$ and the MEC system $k$ on subcarrier $n$, and $N_0$ represents the noise power spectral density of additive white Gaussian noise. 
We assume that any subcarrier of a BS can be allocated to at most one terminal and a terminal can occupy multiple subcarriers. Therefore we have $\sum_{u\in \mathcal{U}} \rho^n_{uk}\leq 1, \forall k,n$.
Therefore, the uplink transmission rate is denoted as 
\begin{equation}
  \label{eq2-3}
  r_u = \sum_{k\in \mathcal{K}}x_{uk}\sum_{n\in \mathcal{N}} \rho^n_{uk}B \log \left( 1 + \it{\Phi}^n_{uk} \right), \forall u.
\end{equation}
The terminal also needs to decide whether to perform semantic extraction and compression of the task based on resource conditions. Let the indicator vector be $\bm{e} = \{e_{ui}\in \{0,1\}\}_{u \in \mathcal{U}, i \in \mathcal{I}}$, where $e_{ui} = 1$ when terminal $u$ perform semantic extraction and compression on task $i$. The uplink transmission delay is expressed as 
\begin{equation}
  \label{eq2-4}
  t^{comm}_u = \frac{\sum\limits_{i\in \mathcal{I}}z_{ui}e_{ui}s_i/\varepsilon_i + \sum\limits_{i\in \mathcal{I}}z_{ui}(1 - e_{ui})s_i}{r_u}, \forall u.
\end{equation}
The inference results of downlink which need to be transmitted to terminals are relatively small compared to the uplink transmission data, therefore it is not considered here.

\subsection{Task Computation Model}
We consider deploying the EEoI mechanism in the deployed model and setting $M$ exit points based on the scenario and task requirements.
Assuming that the exit threshold has been trained to a reasonable value allowing different tasks to exit early with almost the same accuracy\cite{ref3,ref24}, then 
we have $c_i = F_R(s^R_i,m_i),\forall i,\label{eq2-5}$ 
where $F_R(\cdot)$(cycle) represents the relationship between data size, exit point settings and the computation amount required, $s^R_i$ is the input data size in recognition process and $m_i$ represents the exit point of task $i$ while ensuring accuracy.\\
\textbf{Remark 1.} \textit{From the above relationship formula we notice that $F_R(\cdot)$ represents a type of complex relationship between computation amount required and data size. However, this function is usually linear which is proper for traditional computation tasks.
For intelligent scenarios in our model, this linear relationship is no longer suitable for MEC systems to evaluate offloading and resource allocation because intelligent models demand more complicated computation, which will be verified later.
}

According to \cite{ref10}, the computational complexity of convolutional layers is more complicated compared to other types of layer operations (e.g., pooling layers has no parameters and the computational complexity of fully connected layers is input $\times$ output) considering computation of multiple layer in machine learning models which requires special attention.
The floating-point operation (FLOP) counts can be used to measure the computation amount of tasks in hardware, which can characterize computational complexity varies with different amount of convolutional layers. 
When deploying convolutional layer operations using the commonly used matrix multiplication, we can get $2ADC k_w k_h W_{out}{H_{out}}$ as FLOP counts which represents that $A$ feature maps with $C$ channels input to $D$ convolutional filters with shape $k_w \times k_h$, and output $A\times D$ feature maps with shape $W_{out}\times H_{out}$, where the output shape is related to the input shape and the shape of convolutional filters.
Therefore, assuming that input shape is $W_{in}\times H_{in}$, the stride of convolutional filters is set to 1 and the fill length is set to 0, then we have $W_{out} = W_{in} - k_w +1$ and $H_{out} = H_{in} - k_h +1$.

In that way, we decompose the data size of task $i$ and we have $s_i = A_i W_i H_i C_i$. For convenience, we set $|D_{m_i}|$ convolutional layers connected in series before exit point $m_i$, and $L^d$ convolutional filters with shape $k^d_w \times k^d_h$ in each layer where $d \in D_{m_i}$ and $D_{m_i}$ is the set of filters in the process.
There is a fully connected layer before each exit point to classify and output results whose input is $W^R_{i,|D_{m_i|}}\times H^R_{i,|D_{m_i}|}$, i.e., the output size of last convolutional filter. The output of fully connected layers is the result of recognition which is neglected.
Considering the relationship between FLOP counts and cycles, we define $\varPsi = \frac{FLOP}{cycle}$ which represents that floating-point operations per cycle, i.e., the computation hardware can perform $\varPsi$ FLOP counts in one cycle.  
Therefore, the approximation formula for computation amount of inference is given by
  \begin{align}
  \label{eq2-6}
  &F_R(s^R_i,m_i) = \notag \\
  &\sum_{d\in D_{m_i}} 2A_i L^d (W^R_{i,d} - k^d_w + 1) (H^R_{i,d} - k^d_h + 1)C_i k^d_w k^d_h \\
  &{+ A_i C_i W^R_{i,|D_{m_i}|} H^R_{i,|D_{m_i}|}k^R + \sum_{d\in D^R} k^{Rin}_dk^{Rout}_d} \  (\text{FLOP}),\forall i, \notag
  \end{align}
where in recognition process {$s^R_i$}, $W^R_{i,d}$ and $H^R_{i,d}$ represent {the input data size,} the input shape of convolutional filter $d$ respectively, 
{and $k^R$, $D^R$, $k^{Rin}_d$ and $k^{Rout}_d$ are output size of fully connected layer 1, set of fully connected layers excluding the first layer, input and output dimensions of fully connected layers respectively, which depend on network design.}
Note that we have 
$W^R_{i,d+1} = W^R_{i,d}-k^d_w +1$ and 
$H^R_{i,d+1} = H^R_{i,d}-k^d_h +1$ here. \\
\textbf{Remark 2.} \textit{From $\mathrm{(4)}$ we can see that the exit point design is different from traditional inference computation.
{In the case where the exit point threshold has been trained well, tasks have different early exit points during inference to avoid performing complete computation.
Thus, this model can greatly reduce computational overhead.}
Different tasks can be computed with varying degrees of computation complexity if $m_i$ is different. 
In our scenario, this design plays a great role in improving system reward which will be confirmed in Section IV.
} 

According to \cite{ref13}, the process of semantic extraction and compression as well as semantic reconstruction is understood as an end-to-end CNN. The above method is also used to represent the computational complexity of semantic extraction and compression at the terminal and semantic reconstruction at MEC system.
In the process, exit point is not set because its network model is relatively simpler, therefore $m_i$ has no impact on FLOP count. Similarly, the FLOP count for semantic extraction and compression process is denoted as 
\begin{equation}
  \begin{aligned}
  \label{eq2-7}
  F_E(s_i) = &\sum_{d\in D^E_i} 2A_i L^d (W^E_{i,d} - k^d_w + 1)\cdot \\
  &(H^E_{i,d} - k^d_h + 1)C_i k^d_w k^d_h \ (\text{FLOP}),\forall i,
  \end{aligned}
\end{equation}
where $D^E_i$ is the set of convolutional filters and $W^E_{i,d}$, $H^E_{i,d}$ are the input shape of convolutional filter $d$ in semantic extraction and compression process.
We have $W^E_{i,1} = W_i$ and $H^E_{i,1} = H_i$. For semantic reconstruction process, the FLOP count is represented as 
\begin{equation}
  \begin{aligned}
  \label{eq2-8}
  &F_C(s^C_i) = \\
  &\sum_{d\in D^C_i} 2A_i L^d (W^C_{i,d} - k^d_w + 1) (H^C_{i,d} - k^d_h + 1)C_i k^d_w k^d_h \\
  &+{A_i C_i W^C_{i,|D^C_{m_i}|} H^C_{i,|D^C_{m_i}|} k^C + \sum_{d\in D^C}k^{Cin}_d k^{Cout}_d}\  (\text{FLOP}),\forall i,
  \end{aligned}
\end{equation}
where in semantic reconstruction process, $s^C_i$ is the input data size, $D^C_i$ is the set for convolutional filters, $W^C_{i,d}$ and $H^C_{i,d}$ are the input shape of convolutional filter $d$, 
and {$k^C$, $D^C$, $k^{Cin}_d$ and $k^{Cout}_d$ are the output size of fully connected layer 1, set of fully connected layers excluding the first layer, input and output dimensions of fully connected layers respectively.}
Note that the input of semantic reconstruction is feature map which is different from the image shape directly transmitted to MEC systems, therefore the computation cost caused is also different. 
{It can be seen that $s^C_i$ depends on $s_i$ and the network structure of semantic extraction and compression process, and similarly, $s^R_i$ depends on $s^C_i$ and the semantic reconstruction process.}

For convenience, we adopt this type of FLOP count formulas to represent computation amount required to analyze our model. The task computation delay is 
\begin{equation}
  \begin{aligned}
  \label{eq2-9}
  t^{comp}_u = &\frac{\sum\limits_{i\in \mathcal{I}}z_{ui}(1-e_{ui})c_i + \sum\limits_{i\in \mathcal{I}}z_{ui}e_{ui}(F_C(s^C_i)+c_i)}{\varPsi f_u}\\ 
  &+\frac{\sum\limits_{i\in \mathcal{I}}z_{ui}e_{ui}F_E(s_i)}{\varPsi f^L_u},\forall i,
  \end{aligned}
\end{equation}
where $f_u = \sum_{k\in \mathcal{K}} x_{uk}f_{uk},\forall u$, $\bm{f} = \{f_{uk}\geq 0\}_{u\in \mathcal{U},k\in \mathcal{K}}$ is computation capacity allocation vector of terminal $u$ associated with MEC system $k$, and $f^L_u$ is local computation capacity of terminal $u$. 

\section{Problem Formulation and Algorithm Design}
To improve the system performance, we design a reward function based on delay and formulate an optimization problem which is solved using BCD based iteration optimization.
We decompose variables of the problem into two subsets and solve them by 3D matching and convex optimization respectively.
Subsequently, we develop a BCD based iteration algorithm to solve the problem effectively.

\subsection{Problem Formulation}
The design of EEoI mechanism in the image semantic transmission of our model mainly affects the task delay in the MEC system, thereby influencing the system performance.
{To better measure the quality of offloading services in our model, we mainly consider the quality of experience (QoE) based on delay brought by the task offloading of terminals in the system. 
Inspired by the widely used QoE metric, we adopt the formula in \cite{ref25} and make slight changes to make it suitable for intelligent scenarios.}
We define the system reward based on delay and optimize the weighted delay {based on task priority} of terminals. 
The reward of terminal $u$ with task $i$ is denoted as 
\begin{equation}
  \label{eq3-1}
  R_{ui} = C_1 \ln \left( \frac{1}{M_i(t^{comm}_u + t^{comp}_u)} \right) + C_2,\forall u,i,
\end{equation}
where $C_1 > 0$ and $C_2$ is the constant used to adjust the value of system reward.
In that way, we consider optimizing system communication mode $\bm{Z}_1 = \{\bm{x},\bm{\rho}\}$ and computation mode $\bm{Z}_2 = \{\bm{e},\bm{f}\}$ to maximize the weighted delay of all terminals. The optimization problem is represented as 

\begin{subequations}
  \begin{align}
   &\max_{\bm{Z}_1,\bm{Z}_2} \sum_{u\in \mathcal{U}}\sum_{i\in \mathcal{I}} z_{ui}R_{ui} \label{eq3-2}\\
   &\ {\rm{s.t.}}\ x_{uk}\in \{0,1\}, \forall u,k, \label{eq3-2b}\\
   &\ \ \ \ \ \ \rho^n_{uk} \in \{0,1\}, \forall u,k,n, \label{eq3-2c}\\
   &\ \ \ \ \ \ e_{ui}\in\{0,1\}, \forall u,i, \label{eq3-2d}\\  
   &\ \ \ \ \ \ \sum_{k\in \mathcal{K}}x_{uk}\leq 1, \forall u, \label{eq3-2e}\\
   &\ \ \ \ \ \ \sum_{u\in \mathcal{U}} \rho^n_{uk}\leq 1, \forall k,n, \label{eq3-2f}\\
   &\ \ \ \ \ \ \sum_{i\in \mathcal{I}}e_{ui}\leq 1, \forall i,  \label{eq3-2g}\\
   &\ \ \ \ \ \ \sum_{u\in \mathcal{U}}z_{ui}(t^{comm}_u+t^{comp}_u) \leq \tau_i, \forall i, \label{eq3-2h} \\
   &\ \ \ \ \ \ f_{uk}\geq 0,\sum_{u\in \mathcal{U}}x_{uk}f_{uk} \leq F_k, \forall u,k.\label{eq3-2i} 
  \end{align}
\end{subequations}

In \eqref{eq3-2}, the constraints \eqref{eq3-2b}, \eqref{eq3-2c} and \eqref{eq3-2d} ensure that the value of three indicator variables is restrict to 0 and 1. 
Constraint \eqref{eq3-2e} signifies that a terminal can only be associated with one MEC system.
Constraint \eqref{eq3-2f} ensures that one subcarrier $n$ on one MEC system $k$ can only be allocated to one terminal $u$.
Constraint \eqref{eq3-2g} states that a terminal can only have one task to perform semantic extraction and compression.
Constraints \eqref{eq3-2h} and \eqref{eq3-2i} have been proposed to ensure that each task $i$ should remain within its certain delay limits and allocated computation capacity should not exceed the total computation capacity of MEC systems respectively.

\subsection{Algorithm Design}
The above \eqref{eq3-2} is a non-linear mixed integer programming and non-convex optimization problem, which is typically categorized as a NP-hard problem. 
Consequently, we need to decompose it into multiple subproblems and adopt different methods to solve them. 
For the purpose of simplicity in solving \eqref{eq3-2}, we break the variables down into two separate subsets and decompose it into three subproblems by assigning values to additional variables based on BCD to descent iteratively.

\emph{1) Communication part:} In this part, $\bm{Z}_2$ is given in \eqref{eq3-2} to descent $\bm{Z}_1$, i.e., computation variables $\bm{e}$ and $\bm{f}$ are fixed to solve communication variables $\bm{x}$ and $\bm{\rho}$.
Then we can acquire communication resource allocation subproblem which is given as 
\begin{subequations}
  \begin{align}
    &\min_{\bm{x,\rho}} \sum_{u\in \mathcal{U}}\sum_{i\in \mathcal{I}} z_{ui} \ln\left( M_i \left(\frac{A^\alpha_u}{r_u} + \frac{A^\beta_u}{f_u} + A^\gamma_u\right) \right) \label{eq3-3}\\
    &\ {\rm{s.t.}}\ \eqref{eq3-2b},\eqref{eq3-2c},\eqref{eq3-2e},\eqref{eq3-2f} \\
    &\ \ \ \ \ \ \eqref{eq3-2h}': \sum_{u\in \mathcal{U}} z_{ui}\left(\frac{A^\alpha_u}{r_u} + \frac{A^\beta_u}{f_u} + A^\gamma_u\right) \leq \tau_i,\forall i, 
  \end{align} 
\end{subequations}
where $A^\alpha_u = \sum_{i\in \mathcal{I}} z_{ui}e_{ui}s_i/\varepsilon_i + \sum_{i\in \mathcal{I}} z_{ui}(1-e_{ui})s_i$, 
$A^\beta_u = \sum_{i\in \mathcal{I}} z_{ui}(1-e_{ui})c_i + \sum_{i\in \mathcal{I}}z_{ui}e_{ui}\left( F_C(s^C_i)+c_i \right)$, and $A^\gamma_u = \left(\sum_{i\in \mathcal{I}} z_{ui}e_{ui} F_E(s_i)\right)/f^L_u$, which are all constants in this subproblem.
$r_u = \sum_{k\in \mathcal{K}} x_{uk} \sum_{n\in \mathcal{N}}\rho^n_{uk}B\log(1+\phi^n_{uk})$ according to \eqref{eq2-3} and $f_u = \sum_{k\in \mathcal{K}}x_{uk}f_{uk}$.
\eqref{eq3-2h} in \eqref{eq3-2} is converted to \eqref{eq3-2h}$'$ here.

It is obvious that \eqref{eq3-3} is a 3D-matching problem with three sets $(\mathcal{U},\mathcal{K},\mathcal{N})$ and is NP-hard too.
Therefore, we can solve \eqref{eq3-3} by decomposing it into two 2D-matching, i.e., $(\mathcal{U},\mathcal{K})$ and $((\mathcal{U},\mathcal{K}), \mathcal{N})$ to optimize iteratively \cite{ref26}.
Thus, the 2D-matching $(\mathcal{U},\mathcal{K})$ is a many-to-one matching problem obviously.
However, any subcarriers of a BS can be assigned to one terminal at most and a terminal can occupy multiple subcarriers limited by OFDMA.
We assume set {$\varOmega = \{(u,k)|u\in \mathcal{U},k\in \mathcal{K},\sum_{k\in \mathcal{K}}x_{uk}\leq 1\}$} is the solution set of 2D-matching $(\mathcal{U},\mathcal{K})$ and $|\varOmega| = U$.
Therefore, subcarrier allocation can be solved by 2D-matching $(\varOmega, \mathcal{N})$ which is a many-to-many matching problem.

The above proposed matching problems have externalities where each element in the matching set has a dynamic preference list over the opposite set of elements influenced by other elements.
We adopt a preference list over the set of matching states to overcome externalities\cite{ref27}.

\emph{a) Matching Problem Formulation:} 

\textbf{Definition 1 (2D-matching):} a matching $\mu$ is a function from the set $\varUpsilon \cup W$ to the set of all subsets of  $\varUpsilon \cup W$ such that

1) $\mu(y) \subseteq W$ and $|\mu(y)| = l_w$, $\forall y\in \varUpsilon$;

2) $\mu(w) \subseteq \varUpsilon$ and $|\mu(w)| =l_y$, $\forall w\in W$;

3) $\mu(y) \subseteq W$ if and only if $\mu(w)\subseteq \varUpsilon$;

4) $y\in \mu(w)$ if and only if $w\in \mu(y)$;\\
where $\varUpsilon = \{y_1,y_2,...,y_j\}$ and $W = \{w_1,w_2,...,w_q\}$ are two finite and disjoint sets, $l_w$ and $l_y$ are two positive integers. 
Evidently, it is many-to-many matching when $l_w \geq 2$ and $l_y \geq 2$ and is a many-to-one matching when $l_w \geq 2$ and $l_y = 1$. 
{The matching function usually investigates the matching objects and properties of a certain element in a set, so we omit $\mu(\{\cdot\})$ as $\mu(\cdot)$ here when there is only one element in $\{\cdot\}$ and $|\mu(\cdot)|$ indicates the number of elements in the set that match the element.}

Note that this type of matching problems is lack of the property of substitutability\cite{ref26},
and the above matching problems both have externalities\cite{ref27}.
Therefore, given element $y\in \varUpsilon$ has a transitive and strict list, i.e., its interests over the set $W$, and vice versa.
We adopt preference relation symbol $\succ$ to represent the matching relationship for convenience.
Generally, $w_1\succ_y w_2$ represents that element $y$ prefer $w_1$ strictly than $w_2$, and if $w_2\succ_y w_3$ is satisfied we have $w_1\succ_y w_3$.
Due to the existence of externality and non-substitutability, the preference lists of the formulated many-to-many(one) matching problem vary over the matching process which makes the matching mechanism complicated.
Accordingly, we define the swap operation, swap-blocking pair and two-sided stable matching to design our matching algorithms. 

\emph{b) Many-to-one Matching of terminal association:}
In this many-to-one matching problem, we define the preference of terminal $u$ associated with MEC system $k$ as
\begin{equation}
  \label{eq3-4}
  \Phi_{uk}(\mu) = \sum_{i\in \mathcal{I}} z_{ui}\ln\left( M_i \left( \frac{A^\alpha_u}{r_{uk}} + \frac{A^\beta_u}{f_{uk}} +A^\gamma_u\right) \right),\forall u,k,
\end{equation}
where $r_{uk} = \sum_{n\in \mathcal{N}} \rho^n_{uk}B\log(1+\phi^n_{uk})$. For matching $\mu$ and $\mu'$, based on \eqref{eq3-4} we have
\begin{equation}
  \label{eq3-5}
  (k,\mu)\succ_u (k',\mu')\Leftrightarrow \Phi_{uk}(\mu) < \Phi_{uk'}(\mu'),
\end{equation}
which represents that terminal $u$ associated with $k$ has lower total weighted delay compared to being associated with $k'$, i.e., terminal $u$ prefer $k$ in matching $\mu$ rather than $k'$ in matching $\mu'$.
Similarly, the preference of MEC system $k$ associated by subset $\mu(k)$ of $\mathcal{U}$ is denoted as
\begin{equation}
  \label{eq3-6}
  \Phi_{k}(\mu) = \sum_{u\in \mu(k)}\sum_{i\in \mathcal{I}} z_{ui}\ln\left( M_i \left( \frac{A^\alpha_u}{r_{uk}} + \frac{A^\beta_u}{f_{uk}} +A^\gamma_u\right) \right),\forall k.
\end{equation}
In that way, for any two subsets of terminals $\mathcal{U}_1 = \mu(k)$ and $\mathcal{U}_2 = \mu'(k)$ with $\mathcal{U}_1 \neq \mathcal{U}_2$, based on the above formula we have 
\begin{equation}
  \label{eq3-7}
  (\mathcal{U}_1,\mu)\succ_k (\mathcal{U}_2,\mu') \Leftrightarrow \Phi_k(\mu) < \Phi_k(\mu'),
\end{equation}
which means that MEC system $k$ associated by $\mathcal{U}_1$ has lower total weighted delay compared to being associated by $\mathcal{U}_2$, i.e., MEC system $k$ prefer $\mathcal{U}_1$ in matching $\mu$ rather than $\mathcal{U}_2$ in matching $\mu'$.

After the preference lists is formulated, we model the terminal association problem as many-to-one two-sided matching problem based on two discrete sets $\mathcal{U},\mathcal{K}$.  
Due to the externalities, we define the swap operation to ensure the stable matching. For a matching $\mu$, two pairs $(u,k)\in \mu$ and $(u',k')\in \mu$, the swap operation is denoted as
\begin{equation}
  \label{eq3-8}
  \mu^{u'k'}_{uk} = \left\{ \mu\setminus \{(u,k),(u',k')\}\cup \{(u,k'),(u',k)\} \right\},
\end{equation}
where terminals $u$ and $u'$ exchange their matched elements $k$ and $k'$ while keeping all other matching states the same. 
Therefore, a pair $(u,u')$ is a swap-blocking pair if and only if 

1) $\forall j\in \{ u,u',k,k' \}$, we have $\Phi_j(\mu^{u'k'}_{uk})\leq \Phi_j({\mu})$;

2) $\exists j\in \{ u,u',k,k' \}$, we have $\Phi_j(\mu^{u'k'}_{uk})< \Phi_j({\mu})$;\\
where if $j\in \{u,u'\}$, $\Phi_j \in \{\Phi_{uk},\Phi_{u'k},\Phi_{uk'},\Phi_{u'k'}\}$ according to \eqref{eq3-4}.
The aforementioned condition shows 1) that the weighted delay should not increase after the swap operation; 2) that at least the weighted delay of one element decreases after swap operation.
Then the matching $\mu$ is two-sided exchange-stable if and only if there does not exist a swap-blocking pair.

The matching algorithm of terminal association mainly follows the above formulas. 
In the matching initialization phase the random matching is processed according to constraints of \eqref{eq3-3}.
After that, each terminal attempts to search other terminals to find swap-blocking pairs and process the swap operation until there does not exist a swap-blocking pair.
Therefore, the terminal association many-to-one matching algorithm is shown as \textbf{Algorithm \ref{alg:alg1}}.

\begin{algorithm}[H]
  \caption{Terminal Association Many-to-one Matching Algorithm}\label{alg:alg1}
  \begin{algorithmic}[1] 
   \STATE \textbf{Initialize} randomly match the terminal set $\mathcal{U}$ and MEC system set $\mathcal{K}$ that meet the constraints \eqref{eq3-2b}, \eqref{eq3-2c}, \eqref{eq3-2e}, \eqref{eq3-2f} and \eqref{eq3-2h}$'$, and define it as the initial matching state $\mu_1$.
   \STATE \textbf{Swap matching process:}
   \REPEAT
   \STATE For each terminal $u\in \mu_1$, it searches another terminal $u'$ and judge whether $(u,u')$ is a swap-blocking pair;
   \STATE \textbf{if} $(u,u')$ is a swap-blocking pair \textbf{then}
   \STATE $\ \ \mu_1 \leftarrow \mu^{u'k'}_{uk}$; 
   \STATE \textbf{else} 
   \STATE $\ \ $Keep the current matching state;
   \STATE \textbf{end if}
   \UNTIL No swap-blocking pair can be constructed.
   \STATE \textbf{Output} the stable terminal and MEC system matching $\mu^*_1$, association variable $\bm{x^*}$ and the corresponding system reward $\Phi_{L_1} = \Phi(\mu^*_1)$.
  \end{algorithmic}
\end{algorithm}

\emph{c) Many-to-many Matching of subcarrier allocation:} In the many-to-many matching problem, we define the preference formula of the association pair $(u,k)$ (it is expressed as $u$ for simplicity below) in $\varOmega$ over allocated subcarrier $n$ is denoted as
\begin{equation}
  \begin{aligned}
  \label{eq3-9}
  \Phi_{un}(\mu) = &\sum_{i\in \mathcal{I}} z_{ui} \ln\Bigg( M_i \Bigg( \frac{A^\alpha_u}{\sum\limits_{k\in \mathcal{K}}x_{uk}\sum\limits_{j\in \mu_n(u)} B \log(1+\phi^{j}_{uk} )} \\
  &+ \frac{A^\beta_u}{f_{u}} +A^\gamma_u\Bigg) \Bigg),\forall u,n,
  \end{aligned}
\end{equation}
where $\mu_n(u)$ represents that the set of subcarriers allocated to association pair $u$ in the matching when subcarrier $n$ is matched with $u$. For any two matching $\mu$ and $\mu'$, based on \eqref{eq3-9} we have
\begin{equation}
  \label{eq3-10}
  (n,\mu) \succ_u (n',\mu') \Leftrightarrow \Phi_{un}(\mu) < \Phi_{un'}(\mu'),
\end{equation}
which means that association pair $u$ match subcarrier $n$ can decrease the weighted delay compared to $n'$, i.e., association pair $u$ prefer subcarrier $n$ in matching $\mu$ rather than $n'$ in matching $\mu'$.
Similarly, the preference formula of each subcarrier on subset of terminals $\mu(n)$ is represented as
\begin{equation}
  \label{eq3-11}
  \Phi_n(\mu) = \sum_{u\in \mu(n)}\sum_{k\in \mathcal{K}} x_{uk}B \log(1+\phi^n_{uk}),\forall n.
\end{equation}
Note that \eqref{eq3-11} expresses preference through transmission rate because it is meaningless to analyze the weighted delay from the view of each subcarrier.
Therefore, for any two subsets of terminals $\mathcal{U}'_1 = \mu(n)$, $\mathcal{U}'_2 = \mu'(n)$ and $\mathcal{U}'_1 \neq \mathcal{U}'_2$, based on \eqref{eq3-11} we have
\begin{equation}
  \label{eq3-12}
  (\mathcal{U}'_1,\mu) \succ_n (\mathcal{U}'_2,\mu') \Leftrightarrow \Phi_n(\mu) > \Phi_n(\mu'),
\end{equation}
which states that allocating subcarrier $n$ to $\mathcal{U}'_1$ will increase transmission rate compared to $\mathcal{U}'_2$, i.e., subcarrier $n$ prefer $\mathcal{U}'_1$ in the matching $\mu$ rather than $\mathcal{U}'_2$ in the matching $\mu'$.

In this way, we model subcarrier allocation as many-to-many two-sided matching problem based on two discrete sets $\varOmega$ and $\mathcal{N}$.
However, regarding the particularity of subcarriers in OFDMA, the matching need to satisfy the following property: 
we define the terminal set associated with MEC system $k$ and allocated to subcarrier $n$ as $\mathcal{Q}_{kn} = \{u|u\in \mathcal{Q}_k, (u,n)\in \mu\}$, where $\mathcal{Q}_k$ is the terminal set associated with MEC system $k$, and $|\mathcal{Q}_{kn}|\leq 1$, i.e., each subcarrier in MEC system $k$ can only be allocated to at most one terminal.
Since there are unoccupied subcarriers, we denote the hole as $o$ which represents the abstract pair of the unoccupied subcarrier \cite{ref27}.

\begin{algorithm}[H]
  \caption{Subcarrier Allocation Many-to-many Matching Algorithm}\label{alg:alg2}
  \begin{algorithmic}[1] 
   \STATE \textbf{Initialize} randomly match the association pair set $\varOmega$ and subcarrier set $\mathcal{N}$ that meet the constraints \eqref{eq3-2b}, \eqref{eq3-2c}, \eqref{eq3-2e}, \eqref{eq3-2f} and \eqref{eq3-2h}$'$, and define it as the initial matching state $\mu_2$.
   \STATE \textbf{Swap matching process:}
   \REPEAT
   \STATE For each MEC system $k\in \mathcal{K}$: 
   \STATE for each terminal $u\in \mu_2$ associated with $k$, it searches another terminal $u'$ or hole $o$ and judge whether $(u,u')$ or $(u,o)$ is a swap-blocking pair according to \eqref{eq3-14} and \eqref{eq3-15};
   \STATE \textbf{if} $(u,u')$ or $(u,o)$ is a swap-blocking pair \textbf{then}
   \STATE $\ \ \mu_2 \leftarrow \mu^{u'n'}_{un}$ or $\mu_2 \leftarrow \mu^{u'o}_{un}$; 
   \STATE \textbf{else} 
   \STATE $\ \ $Keep the current matching state;
   \STATE \textbf{end if}
   \UNTIL No swap-blocking pair can be constructed.
   \STATE \textbf{Output} the stable association pair and MEC system matching $\mu^*_2$, subcarrier allocation variable $\bm{\rho^*}$ and the corresponding system reward $\Phi_{L_2} = \Phi(\mu^*_2)$.
  \end{algorithmic}
\end{algorithm}

Similarly, due to the externalities of the many-to-many matching problem, we define the swap operation to ensure the stable matching, i.e., allowing that each terminal associated with MEC system exchange subcarriers is necessary. 
For a matching $\mu$ with $n\in \mu(u)$, $n'\in \mu(u')$, $n\notin \mu(u')$, $n'\notin \mu(u)$, the swap operation is defined as
\begin{equation}
  \begin{aligned}
  \label{eq3-13}
    &\mu^{u'n'}_{un} = \big\{ \mu \backslash \{ (u,\mu(u)),(u',\mu(u')) \} \big\} \cup \\
    &\big\{ (u,\{ \{ \mu(u)\backslash \{n\} \}\cup\{n'\} \}),(u',\{ \{ \mu(u')\backslash\{n'\} \} \cup \{n\} \} ) \big\},
  \end{aligned}
\end{equation}
which represents that a swap operation ensures that associated terminals $u$ and $u'$ exchange their matched subcarriers and the matching state is stable simultaneously.
In the matching, a pair $(u,u')$ is swap-blocking pair if and only if

1) the following conditions are all established:
\begin{equation}
  \begin{aligned}
  \label{eq3-14}
    &\Phi_{un'}(\mu^{u'n'}_{un}) \leq \Phi_{un}(\mu), \Phi_{u'n}(\mu^{u'n'}_{un}) \leq \Phi_{u'n'}(\mu),\\
    &\Phi_n(\mu^{u'n'}_{un}) \geq \Phi_n(\mu), \Phi_{n'}(\mu^{u'n'}_{un}) \geq \Phi_{n'}(\mu);\\
  \end{aligned}
\end{equation}

2) at least one of the following conditions is established:
\begin{equation}
  \begin{aligned}
  \label{eq3-15}
    &\Phi_{un'}(\mu^{u'n'}_{un}) < \Phi_{un}(\mu),\Phi_{u'n}(\mu^{u'n'}_{un}) < \Phi_{u'n'}(\mu),\\
    &\Phi_n(\mu^{u'n'}_{un}) > \Phi_n(\mu), \Phi_{n'}(\mu^{u'n'}_{un}) > \Phi_{n'}(\mu),\\
  \end{aligned}
\end{equation}
where 1) represents the weighted delay does not increase and transmission rate does not decrease after swap operation, and 2) represents that the weighted delay of at least one terminal decreases and the transmission rate of at least one terminal increases.
To sum up, the subcarrier allocation many-to-many matching algorithm is shown as \textbf{Algorithm \ref{alg:alg2}}.

\emph{2) Semantic transmission and computation part:} In this part, $\bm{Z}_1$ is given in \eqref{eq3-2} to descent $\bm{Z}_2$, i.e., communication variables $\bm{x}$ and $\bm{\rho}$ are fixed to solve computation variables $\bm{e}$ and $\bm{f}$.
Then we can acquire computation subproblem which is given as 

\begin{subequations}
  \begin{align}
    &\min_{\bm{e},\bm{f}} \sum\limits_{u\in \mathcal{U}}\sum\limits_{i\in \mathcal{I}}z_{ui}ln\Bigg( M_i \Bigg( \frac{\sum\limits_{i\in \mathcal{I}} z_{ui}e_{ui}(s_i/\varepsilon_i - s_i)}{r_u} + \nonumber \\
    &\frac{\sum\limits_{i\in \mathcal{I}}z_{ui}e_{ui}F_C(s_i^C) + \sum\limits_{i\in \mathcal{I}}z_{ui}c_i}{f_u} + \frac{\sum\limits_{i\in \mathcal{I}}z_{ui}e_{ui}F_E(s_i)}{f^L_u} +\beta_u \Bigg)\Bigg) \label{eq3-16} \\
    &\ {\rm{s.t.}} \ \eqref{eq3-2d},\eqref{eq3-2g},\eqref{eq3-2h},\eqref{eq3-2i},
  \end{align}
\end{subequations}
where $\beta_u = \sum_{i\in \mathcal{I}}z_{ui}s_i/ r_u$ is a constant.
The problem \eqref{eq3-16} is still complex with variables of different properties. It is non-convex and still a mixed integer optimization problem.
Therefore, we decompose the two integer and real variables into two subproblems because of the physical meaning of $\bm{e}$ which can not be relaxed into real domain. 

We have semantic transmission subproblem after fixing $\bm{f}$ which is denoted as
\begin{subequations}
  \begin{align}
     &\min_{\bm{e}} \sum\limits_{u\in \mathcal{U}}\sum\limits_{i\in \mathcal{I}}z_{ui}ln\Bigg( M_i \Bigg( \frac{\sum\limits_{i\in \mathcal{I}} z_{ui}e_{ui}(s_i/\varepsilon_i - s_i)}{r_u} +\nonumber \\
     &\frac{\sum\limits_{i\in \mathcal{I}}z_{ui}e_{ui}F_C(s_i^C)}{f_u} + \frac{\sum\limits_{i\in \mathcal{I}}z_{ui}e_{ui}F_E(s_i)}{f^L_u} +B^{\alpha}_u \Bigg)\Bigg) \label{eq3-17} \\
     &\ {\rm{s.t.}} \ \eqref{eq3-2d},\eqref{eq3-2g},\eqref{eq3-2h},
  \end{align}
\end{subequations}
where $B^{\alpha}_u = \sum_{i\in \mathcal{I}}z_{ui}s_i/ r_u+ \sum_{i\in \mathcal{I}}z_{ui}c_i/ f_u$ is a constant in the subproblem.
\begin{figure*}
  \begin{equation}
      \label{eq3-18}
      e^*_{ui} = \left\{ \begin{aligned}
      &\mathop{\arg\min}\limits_{e_{ui}\in\{0,1\}} z_{ui}ln\Bigg( M_i \Bigg( \frac{\sum\limits_{i\in \mathcal{I}} z_{ui}e_{ui}(s_i/\varepsilon_i - s_i)}{r_u} + \frac{\sum\limits_{i\in \mathcal{I}}z_{ui}e_{ui}F_C(s_i^C)}{f_u} + \frac{\sum\limits_{i\in \mathcal{I}}z_{ui}e_{ui}F_E(s_i)}{f^L_u} +B^{\alpha}_u \Bigg)\Bigg), &if z_{ui} = 1,\\
      &0, &if z_{ui} = 0.
      \end{aligned}\right.
  \end{equation}
  \hrule
\end{figure*}  
It is obviously a binary discrete problem and the optimal solution is expressed as \eqref{eq3-18}.

We can get computation capacity subproblem by fixing $\bm{e}$ which is denoted as 
\begin{subequations}
  \begin{align}
     &\min_{\bm{f}} \sum\limits_{u\in \mathcal{U}}\sum\limits_{i\in \mathcal{I}}z_{ui}ln\Bigg( M_i \Bigg(C^\alpha_u + \nonumber \\
     &\frac{\sum\limits_{i\in \mathcal{I}}z_{ui}(1-e_{ui})c_i + \sum\limits_{i\in \mathcal{I}}z_{ui}e_{ui}(F_C(s^C_i) + c_i)}{f_u} \Bigg)\Bigg) \label{eq3-19} \\
     &\ {\rm{s.t.}} \ \eqref{eq3-2h},\eqref{eq3-2i},
  \end{align}
\end{subequations}
where $C^\alpha_u = \left(\sum\limits_{i\in \mathcal{I}}z_{ui}e_{ui}s_i/\varepsilon_i + \sum\limits_{i\in \mathcal{I}}z_{ui}(1-e_{ui})s_i\right)\bigg/r_u + \sum\limits_{i\in \mathcal{I}}z_{ui}e_{ui}F_E(s_i)/f^L_u$ is a constant in the subproblem.

Evidently, the constant terms are greater than 0 and \eqref{eq3-2h}, \eqref{eq3-2i} are affine when $z_{ui} = 1$, therefore \eqref{eq3-19} is convex.
We adopt CVX tools to solve the convex problem through convex optimization \cite{ref28,ref29}.
In that way, the computation subproblem is solved through iteration of \eqref{eq3-17} and \eqref{eq3-19} and the algorithm is summarized as \textbf{Algorithm \ref{alg:alg3}}.
\begin{algorithm}[H]
  \caption{Semantic Transmission and Computation Capacity Allocation Iterative Algorithm}\label{alg:alg3}
  \begin{algorithmic}[1] 
   \STATE \textbf{Initialize} 
   \STATE Obtain the value of $\bm{x}$ and $\bm{\rho}$, and set the initial value of $\bm{f}>0$;
   \STATE Set the iteration count $L_{A3} = 0$, the value of \eqref{eq3-16} $V_{L_{A3}} = 0$, the iteration constraint $\zeta > 0$ and the maximum number of iterations $L^{max}_{A3}$
   \STATE \textbf{Iteration process:}
   \REPEAT
   \STATE For all $u\in \mathcal{U}$, obtain $\bm{e}^{L_{A3}}_{ui}$ according to \eqref{eq3-17} through \eqref{eq3-18};
   \STATE For all $u\in \mathcal{U}$, obtain $\bm{f}^{L_{A3}}$ according to \eqref{eq3-19} through convex optimization;
   \STATE Update $L_{A3} = L_{A3}+1$.
   \STATE Update the value of $V_{L_{A3}}$ through $\bm{e}^{L_{A3}}_{ui}$ and $\bm{f}^{L_{A3}}$;
   \UNTIL $|V_{L_{A3}} - V_{L_{A3}-1}|\leq \zeta$ or $L_{A3} > L^{max}_{A3}$.
   \STATE \textbf{Output} the solution $\bm{e}^*$ and $\bm{f}^*$.
  \end{algorithmic}
\end{algorithm}

\begin{algorithm}[H]
  \caption{BCD Based Joint Semantic Transmission and Resource Allocation Algorithm}\label{alg:alg4}
  \begin{algorithmic}[1] 
   \STATE \textbf{Initialize} 
   \STATE Set the initial feasible value $(\bm{x}^{(0)},\bm{\rho}^{(0)},\bm{e}^{(0)},\bm{f}^{(0)})$.
   \STATE Set the iteration count $L_{A} = 0$, the value of \eqref{eq3-2} $V^{(0)} = 0$, the iteration constraint $\epsilon > 0$ and the maximum number of descent $L^{max}$. 
   \STATE \textbf{Iteration process:}
   \REPEAT
   \STATE \textbf{Step 1:} Terminal association 
   \STATE With given $(\bm{\rho}^{(L_A)},\bm{e}^{(L_A)},\bm{f}^{(L_A)})$, obtain $\bm{x}^{(L_A+1)}$ via \textbf{Algorithm 1} to descent $\bm{Z}_1$;
   \STATE \textbf{Step 2:} Subcarrier allocation 
   \STATE With given $(\bm{x}^{(L_A+1)},\bm{e}^{(L_A)},\bm{f}^{(L_A)})$, obtain $\bm{\rho}^{(L_A+1)}$ via \textbf{Algorithm 2} to descent $\bm{Z}_1$;
   \STATE \textbf{Step 3:} Semantic transmission and computation capacity allocation
   \STATE With given $(\bm{x}^{(L_A+1)},\bm{\rho}^{(L_A+1)})$, obtain $(\bm{e}^{(L_A+1)}$, $\bm{f}^{(L_A+1)})$ via \textbf{Algorithm 3} to descent $\bm{Z}_2$;
   \STATE Update $L_{A} = L_{A}+1$ and the value of $V^{(L_A)}$;
   \UNTIL $|V^{(L_{A})} - V^{(L_{A}-1)}|\leq \epsilon$ or $L_{A} > L^{max}$;
   \STATE \textbf{Output} the solution $(\bm{x}^{(L_A)},\bm{\rho}^{(L_A)},\bm{e}^{(L_A)},\bm{f}^{(L_A)})$.
  \end{algorithmic}
\end{algorithm}

\subsection{Algorithm Overview and Analysis}
The overall algorithm for \eqref{eq3-2} is summarized as \textbf{Algorithm \ref{alg:alg4}} based on BCD which is composed of variable subset $\bm{Z}_1$ with 3D-maching communication part and $\bm{Z}_2$ with semantic transmission and computation part. 

We analyze the convergence and complexity of two 2D-matching algorithms first based on following propositions including stability, convergence and complexity.

1) \textit{Proposition 1 (Stability):  The final matching $\mu^*_1$ and $\mu^*_2$ in \textbf{Algorithm \ref{alg:alg1}} and \textbf{Algorithm \ref{alg:alg2}} respectively are both two-sided exchange-stable matching.}

\textit{Proof:} This proposition can be proved by contradiction. In \textbf{Algorithm \ref{alg:alg1}}, Assume that there exist a blocking pair $(u,u')$ in the final matching $\mu^*_1$ satisfying that $\forall j\in \{u,u',k,k'\}, \Phi_j(\mu^{u'k'}_{uk}) \leq \Phi_j(\mu)$ and $\exists j\in \{u,u',k,k'\}, \Phi_j(\mu^{u'k'}_{uk})<\Phi_j(\mu)$.
The swap operation will continue because the swap-blocking pair is found according to step 2 to step 10, i.e., $\mu^*_1$ is not the final matching state, which contradicts the initial assumption and the proposition is proved.
Therefore, the proposed algorithm reaches a two-sided exchange stability in the end. 
The proof of $\mu^*_2$ in \textbf{Algorithm \ref{alg:alg2}} can be derived similarly which is neglected here for brevity.

2) \textit{Proposition 2 (Convergence): Both \textbf{Algorithm \ref{alg:alg1}} and \textbf{Algorithm \ref{alg:alg2}} converge to a two-sided exchange-stable matching in a finite iterations.}

\textit{Proof:} The convergence of algorithm depends on swap operations. In \textbf{Algorithm \ref{alg:alg1}}, the weighted delay of at least a MEC system $k$ or $k'$ decrease after the relative swap operation according to \eqref{eq3-8}.
Thus, there exists three cases: i) $\Phi_k(\mu')<\Phi_k(\mu)$ and $\Phi_{k'}(\mu')<\Phi_{k'}(\mu)$; ii) $\Phi_k(\mu')=\Phi_k(\mu)$ and $\Phi_{k'}(\mu')<\Phi_{k'}(\mu)$; iii) $\Phi_k(\mu')<\Phi_k(\mu)$ and $\Phi_{k'}(\mu')=\Phi_{k'}(\mu)$. 
It can be observed that the reward of involved MEC systems are non-decreasing and the achievable weighted delay of each MEC system has a lower bound limited by communication resource constraints in practice.
Therefore, the number of iterations of \textbf{Algorithm \ref{alg:alg1}} is finite and it converges to a two-sided exchange-stable matching after there are no swap-blocking pairs.
The convergence proof for \textbf{Algorithm \ref{alg:alg2}} can be derived similarly which is neglected here for brevity.

3) \textit{Proposition 3 (Complexity): The complexity of \textbf{Algorithm \ref{alg:alg1}} and \textbf{Algorithm \ref{alg:alg2}} is upper bounded by $\mathcal{O}(UK+U^{max}KL_{A1})$ and $\mathcal{O}(UKN(1+UNL_{A2}))$ respectively.}

\textit{Proof:} The main part of complexity of matching algorithm is from initialization and swap operation process.
In \textbf{Algorithm \ref{alg:alg1}}, initialization process randomly match sets $\mathcal{U}$ and $\mathcal{K}$ to construct initial pairs where random matching is performed according to resource constraints of each MEC system, and the complexity is $\mathcal{O}(UK)$ in the worst case.
In swap matching process, assume that at most $U^{max}$ terminals process swap operation with other $K-1$ unassociated MEC systems and the total number of iterations is $L_{A1}$, then the complexity is $\mathcal{O}(U^{max}KL_{A1})$. Thus the complexity of \textbf{Algorithm \ref{alg:alg1}} is $\mathcal{O}(UK+U^{max}KL_{A1})$.
Similarly, in \textbf{Algorithm \ref{alg:alg2}}, the complexity of initialization process is $\mathcal{O}(UKN)$ performing matching between association pairs and subcarriers. In swap operation process, each association pair performs swap operation with other subcarriers and check swap-blocking pairs, thus the complexity is $\mathcal{O}(UK(U-1)N(N-1)L_{A2})$.
Therefore, the overall complexity is $\mathcal{O}(UKN(1+UNL_{A2}))$.

Then for \textbf{Algorithm \ref{alg:alg3}}, we assume that the weighted delay value of iteration $L_{A3}$ is $V_{L_{A3}} = V(\bm{Z_1},\bm{e}^{L_{A3}},\bm{f}^{L_{A3}})$. We have
$V(\bm{Z_1},\bm{e}^{L_{A3}},\bm{f}^{L_{A3}}) \geq V(\bm{Z_1},\bm{e}^{L_{A3}+1},\bm{f}^{L_{A3}})$ and 
$V(\bm{Z_1},\bm{e}^{L_{A3}+1},\bm{f}^{L_{A3}}) \geq V(\bm{Z_1},\bm{e}^{L_{A3}+1},\bm{f}^{L_{A3+1}})$ 
after the optimization solution of step 6 and step 7. Thus the weighted delay value is non-increasing after iterations, i.e., $V(\bm{Z_1},\bm{e}^{L_{A3}},\bm{f}^{L_{A3}}) \geq V(\bm{Z_1},\bm{e}^{L_{A3}+1},\bm{f}^{L_{A3+1}})$ always holds, which represents that the algorithm converges after finite number of iterations limited by resource constraints.
Therefore, the overall complexity is $\mathcal{O}((U+(UK)^{3.5})L_{A3})$ while performing iterative computation.

Based on the analysis above, the convergence of \textbf{Algorithm \ref{alg:alg4}} is guaranteed as long as $L_{max}$ is set large enough.
The complexity of \textbf{Algorithm \ref{alg:alg4}} is composed of \textbf{Algorithm \ref{alg:alg1}-\ref{alg:alg3}}, i.e., $\mathcal{O}_1$, $\mathcal{O}_2$ and $\mathcal{O}_3$ respectively, which is denoted as $\mathcal{O} = L_A(\mathcal{O}_1 + \mathcal{O}_2 + \mathcal{O}_3)$.
The overall complexity is $\mathcal{O}(L_A(UK+U^{max}KL_{A1}+UKN(1+UNL_{A2})+(U+(UK)^{3.5})L_{A3}))$.

\section{Simulation Result}
In this section, {we first deploy a DNN pipeline of semantic transmission and EEoI to verify the feasiblity of our model.
Then, }we demonstrate our channel model and simulation parameter design. 
Subsequently, our simulation findings are presented to assess the efficacy of our suggested algorithms.

{
\subsection{Deployment and Comparison}
We adopt semantic transmission network from \cite{ref36} and EEoI from \cite{ref3} to deploy the model we mentioned in Section II, which are recent efficient semantic communication network and classice EEoI network respectively.
The semantic transmission network, named DeepJSCC-V, consists of semantic extraction and compression part and semantic reconstruction part, both of which contain 5 convolutional layers and 5 fully connected layers.
The EEoI architecture adopts B-Alexnet, which adds two exit point designs based on 5 convolutional layers and 3 fully connected layers of Alexnet.
The first exit point adds 2 convolutional layers and 1 fully connected layer after the first convolutional layer of the main network, and the second exit point adds 1 convolutional layer and 2 fully connected layers after the second convolutional layer of the main network.
}

{
To compare the actual delay with the estimated delay of our model in Section II, we adopt CIFAR-10 dataset to train the pipeline and measure the delay and performance of inference as shown in Table II.
We measure average delay of test set in CIFAR-10 on NVIDIA GeForce GTX 4090 24G GPU and design three types of threshold on EEoI. 
Note that the hardware usage of the two networks is different when running inference due to the software architecture.
The estimated delay is very close to the actual delay, which verifies the effectiveness of our FLOP counts model in Section II.
The performance is shown in the last column of the table, i.e., similarity/accuracy, which indicates the DeepJSCC-V and EEoI networks are useful.
{Meanwhile, it is evident that the initial threshold of 0.0001 corresponds to an exit rate of $18.68\%$ at the first exit point. As the second threshold increases, the exit rate at the second exit point rises from $27.36\%$ to $37.59\%$, consequently leading to a reduction in the proportion of tasks that exit at the final point. It is noteworthy that as the second threshold increases, the delay decreases while the accuracy experiences a slight improvement. This phenomenon occurs because the accuracy at each exit point exhibits an extremum in relation to the entropy variation of the task set~\cite{ref3}, thereby illustrating the inherent trade-off between accuracy and latency characteristic of the EEoI framework.}
}
\begin{table}[!t]
  \renewcommand\arraystretch{1.5}
  \fontsize{8}{7}\selectfont
  \caption{{Deployment Performance of Networks}}
  \centering
  \begin{tabular}{p{1.6cm}<{\centering}|p{1.1cm}<{\centering}|p{0.9cm}<{\centering}|p{1.0cm}<{\centering}|p{1.0cm}<{\centering}|p{1.0cm}<{\centering}}
  \hline
  \hline
  {\textbf{Network}} & {\textbf{Threshold}} & {\textbf{Exit(\%)}} & {\textbf{Time(ms)}} & {\textbf{Esm. T(ms)}} & {\textbf{Sim./Acc. (\%)}}\\
  \hline
  DeepJSCC-V & - & - & 0.3613 & 0.3762 & 98.83 \\
  \hline
  B-Alexnet & \{0.0001, 0.001\} & \{18.68, 27.36, 53.96\} & 2.6110 & 2.6583 & 76.26 \\
  \hline 
  B-Alexnet & \{0.0001, 0.005\} & \{18.68, 34.24, 47.08\} & 2.4684 & 2.3206 & 76.29 \\
  \hline
  B-Alexnet & \{0.0001, 0.01\} & \{18.68, 37.59, 43.73\} & 2.3968 & 2.1562 & 76.75 \\
  \hline
  \hline
  \end{tabular}
\end{table}

\subsection{Simulation Parameters}
We examine the simulation of uplink transmission at the system level in a $500m \times 500m$ small cell heterogeneous cellular scenario with four small BSs deployed MEC servers providing association and computation capacity for terminals.
In this area, 20 terminals from $\mathcal{U}$ with 15 types of tasks from $\mathcal{I}$ are active and prepare for processing image recognition.
Considering the scenario and complexity, we adopt channel model from \cite{ref30} which the pass loss from MEC system $k$ to terminal $u$ is denoted as 
\begin{equation}
  \label{eq4-1}
  g_{uk}[dB] = 42.6 + 26 \lg (d_{uk}[km]) + 20 \lg(F^q[MHz]),\forall u,k,
\end{equation}
where $d_{uk}$ is distance between $u$ and $k$, $F^q$ represents the carrier frequency. 
Besides, we set the shadowing which corresponds to normal distribution $N(0,8)$.
We set abstract computation force supply of computation hardware involved with intelligent computing of neural network, i.e., number of stream processors multiple core frequency of GPU\cite{ref20}. 
The employed system parameters are shown as TABLE III.

\begin{table}[!t]
  \renewcommand\arraystretch{1.5}
  \fontsize{8}{7}\selectfont
  \caption{Simulation Parameters}
  \centering
  \begin{tabular}{p{4.8cm}<{\centering}|p{2.1cm}<{\centering}}
  \hline
  \hline
  {\textbf{Parameter}} & {\textbf{Value}} \\
  \hline
  Number of users, $U$ & 20\\
  \hline
  Number of tasks, $I$ & 15\\
  \hline
  Number of subcarriers, $N$ & 32\\
  \hline
  Transmit power, $P^n_{uk}$ & 0.1 W \\
  \hline
  Noise power spectral density, $N_0$ & -174 dBm/Hz \\
  \hline
  Total Bandwidth, $B\times N$ & 32 MHz \\
  \hline
  Carrier frequency, $F^q$ & 2.4 GHz \\
  \hline
  Computing capacity of edge servers, $F_k$ & 2048*0.96 Gigacycle/s \\
  \hline
  Computing capacity of local device, $f^L_u$ & 256*0.96 Gigacycle/s \\
  \hline
  Floating point operations per cycle, $\varPsi $ & 8 FLOP/cycle\\
  \hline
  \hline
  \end{tabular}
\end{table}

Considering the task parameters, we set the input as a square feature map with three channels.
We also assume that the deployed CNN has been adjusted to input a feature map with the shape of uniform distribution values restricted by upper and lower bounds to correspond to the setting of task data size range \cite{ref13,ref19}.
The convolution filters are set as $3\times 3$ and $5\times 5$ which are generally adopted in EEoI and three exit points are trained to suitable states in final recognition process.
Number of convolution layers before exit points is integer within $[3,5]$. The delay limits and priorities of different tasks are taken as integer values using a random distribution method.
Thus, task parameters is represented in TABLE IV \cite{ref3,ref13}.
\begin{table}[!t]
  \renewcommand\arraystretch{1.5}
  \fontsize{9}{7}\selectfont
  \caption{Task Parameters}
  \centering
  \begin{tabular}{p{5.1cm}<{\centering}|p{1.8cm}<{\centering}}
    \hline
    \hline
    {\textbf{Parameter}} & {\textbf{Value}}\\
    \hline
    Shape of feature map $W_i,H_i$& $[224,448]$ \\ 
    \hline
    Shape of convolution filters $k^d_w,k^d_h$ & ${3,5}$\\
    \hline
    Compression ratio $\varepsilon_i$ & $(1,5]$ \\
    \hline
    Number of feature map, channels and convolution filters $A_i,C_i,D$ & $1,3,96$ \\
    \hline
    Delay constraint $\tau_i$ & $[0.5,1]$s \\
    \hline
    Number of convolution layers in recognition process $|D_{m_i}|$ & ${3,4,5}$ \\
    \hline
    Number of convolution layers in other process $|D^E_i|,|D^C_i|$ & $4,1$ \\
    \hline
    \hline
  \end{tabular}
\end{table}

\subsection{Performance of the Proposed Algorithm}
In order to validate the effectiveness of the proposed algorithm, we present the following comparison schemes as a means of assessing its performance:
\begin{itemize}
\item{\textbf{Linear Computation Model (Linear)}\cite{ref19}: The computation amount required of each type of task is modelled as linear function, i.e., we calculate the average computation density $\theta_{m_i}$ through the task computation amount required of each exit point.
{In addition, we need to divide it into several groups according to the number of exit points with different computation cost,} which is denoted as
$\theta_{m_i} = \frac{\sum_{i\in M^S_{m_i}}c_i/s_i}{|M^S_{m_i}|},\forall i\in \mathcal{I}, $
where $M^S_{m_i}$ is the task set with exit point $m_i${, and the numerator is the sum of task computation density with exit point $m_i$.}
}
\item{\textbf{Fixed Transmission Mode (FTM)}: The scheme adopts given transmission mode $\bm{e}$ consistent with uniform distribution.}
\item{\textbf{Fixed Association (FA)}: The scheme entails fixed association mode $\bm{x}$ based on minimum distance.}
\item{\textbf{Uniform Computation (UC)}: The scheme distributes edge computation capacity $\bm{f}$ evenly.}
\end{itemize}

\begin{figure}[!t]
  \centering
  \includegraphics[scale = 0.6]{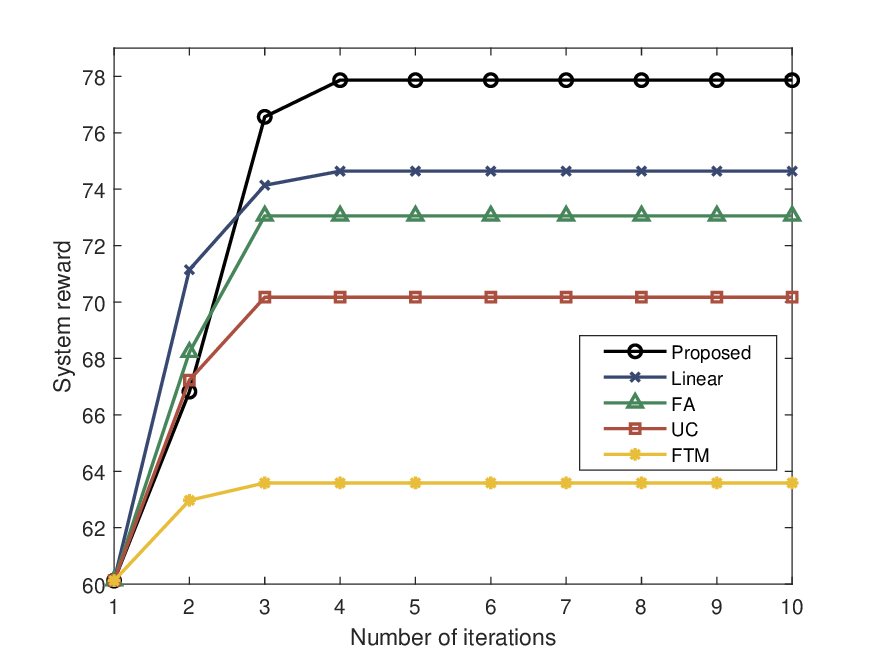}
  \caption{{Convergence of all algorithms.}}
  \label{fig_2}
\end{figure}
In Fig. \ref{fig_2}, we illustrate the convergence behavior of all schemes. 
All algorithms convergence at an accelerated pace and the convergence times of different algorithms are slightly different.
The difference mainly comes from the different optimize performance and the overall convergence performance is good.
The proposed algorithm has a better optimization effect on system reward with a {$4.3\%$} improvement compared to the Linear algorithm under the default parameters.
Note that for comparison purposes here, linear computation model uses the average of the computation amount required of all tasks to obtain the computation density $\theta_{m_i}$. 
Although the performance has declined here, it is still a relatively accurate estimate for the system. 
However, it is difficult to first find the average value of the computation force cost of a certain type of task to represent the linear computing density value in practice. 
In most cases, an empirical approximation is used so that the effect may be worse.
The proposed algorithm has {$6.6\%$, $11.0\%$ and $22.5\%$} improvements compared to the FA algorithm, the UC algorithm and the FTM algorithm respectively. 
{The proposed algorithm has better optimization performance compared with the benchmarks and an acceptable convergence speed, i.e., it improves performance without increasing the complexity of the algorithm significantly.}
It is better adapted to the MEC system than linear computation model in our scenario which is a good proof of \textbf{Remark 1}.

\begin{figure}[!t]
  \centering
  \includegraphics[scale = 0.6]{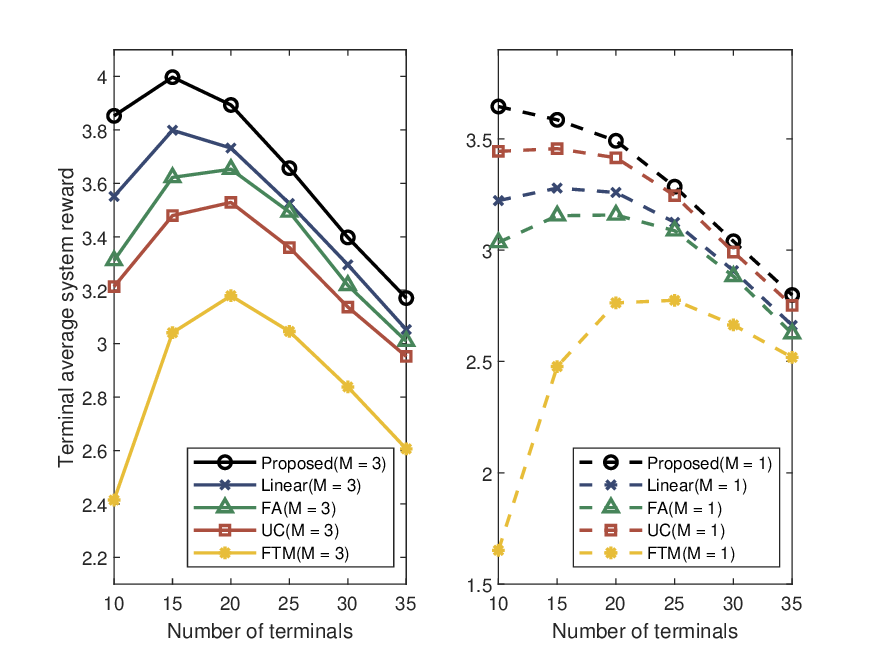}
  \caption{{Terminal average system reward with varying number of terminals under 3 exit points and 1 exit point.}}
  \label{fig_3}
\end{figure}
The terminal average system reward with varying number of terminals $U$ is examined under different number of exit points, i.e., $M = 1$ (i.e., no exit point design) and $M = 3$, as illustrates in Fig. \ref{fig_3}.
{Note that we illustrate the terminal average system reward here to present the average terminal performance of the system in a more intuitive way than system reward varying with number of terminals.} 
From the left subfigure, the proposed algorithm always has a performance gain in terms of average terminal reward compared to other algorithms while the number of terminals changing.
Compared with the Linear algorithm, FA algorithm, UC algorithm, and FTM algorithm, the proposed algorithm has average performance improvements of {$4.8\%$, $8.2\%$, $11.6\%$, and $29.2\%$} respectively.
Comparing the two subfigures on left and right, the performance improvement of the proposed algorithm is {$10.9\%$} on average when number of exit point $M$ is different. 
The impact of the exit point is greater than that of the Linear algorithm when the number of terminals changes.
At the same time, terminal average system reward trends of different algorithms are similar when $M = 3$ and the maximum value is obtained when $U = 15$ or $20$.
As the number of terminals becomes larger, the terminal average system reward decreases and approaches each other. 
The trend is similar with that when $M = 1$ but the maximum value point moves forward and system reward is smaller.
This is because resources are not fully utilized when the number of terminals is low.
The algorithm with better performance has a closer maximum value point to the front because it can make fuller use of resources when the number of users is lower.
However, when the number of terminals is relatively large, resource competition is fierce and the difference in terminal average system reward of the algorithms decreases.
The task computation force overhead of the system is lower when $M = 3$ and the resource consumption is relatively large when $M = 1$. 
It is easier to fully utilize the system resource when the number of terminals is low and the maximum value point is closer to the front. 
Overall, the proposed algorithm has better performance than comparison algorithms with varying number of terminals in this system scenario. 
In comparison, the design of early exit points has a greater impact on system performance which reflects the necessity of exit point design in this scenario and verifies \textbf{Remark 2}.

\begin{figure}[!t]
  \centering
  \includegraphics[scale = 0.6]{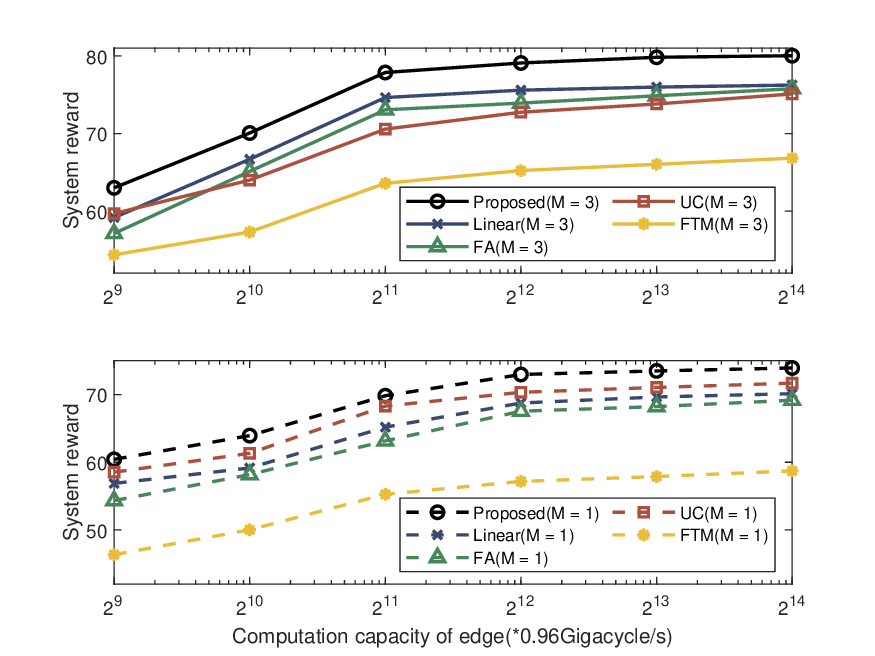}
  \caption{{{System reward with varying computing capacity of edge under 3 exit points and 1 exit point.}}}
  \label{fig_4}
\end{figure}
In Fig. \ref{fig_4}, we compare system reward with computing capacity $F_k$ of edge under different number of exit points. 
Notably, the x-axis values here are in exponential form and the logarithmic abscissa is used to represent changes in computation capacity of edge. 
From the upper subfigure, compared with the Linear algorithm, FA algorithm, UC algorithm, and FTM algorithm, the proposed algorithm has average performance improvements of {$5.1\%$, $7.3\%$, $8.1\%$, and $20.4\%$} respectively.
Comparing the upper and lower subfigures, the performance improvement of the proposed algorithm is {$8.4\%$} on average when number of exit point $M$ is different. 
At the same time, both figures show that as computation capacity of edge increases, the system reward increases and then becomes stable. 
This is because after computation capacity of edge increases to a certain amount, it is limited by other resources and cannot continue to increase system reward resulting in limited performance.
When $M = 3$, the UC algorithm changes more gently and the performance is least affected when $M$ changes. 
However, due to the increase in task computation force overhead when $M = 1$, the system reward decreases and the trend of each algorithm becomes smoother than that when $M = 3$ except for the UC algorithm which evenly allocates computation capacity.
When $M = 1$, computation capacity of edge is relatively more limited and the performance gain is smaller while optimizing computation capacity, thus showing the UC algorithm is least affected by changes in number of exit points.
In short, the proposed algorithm has better performance than the comparison schemes with varying computation capacity of edge and optimizing that can achieve better performance when there are multiple exit points in the design.

\begin{figure}[!t]
  \centering
  \includegraphics[scale = 0.6]{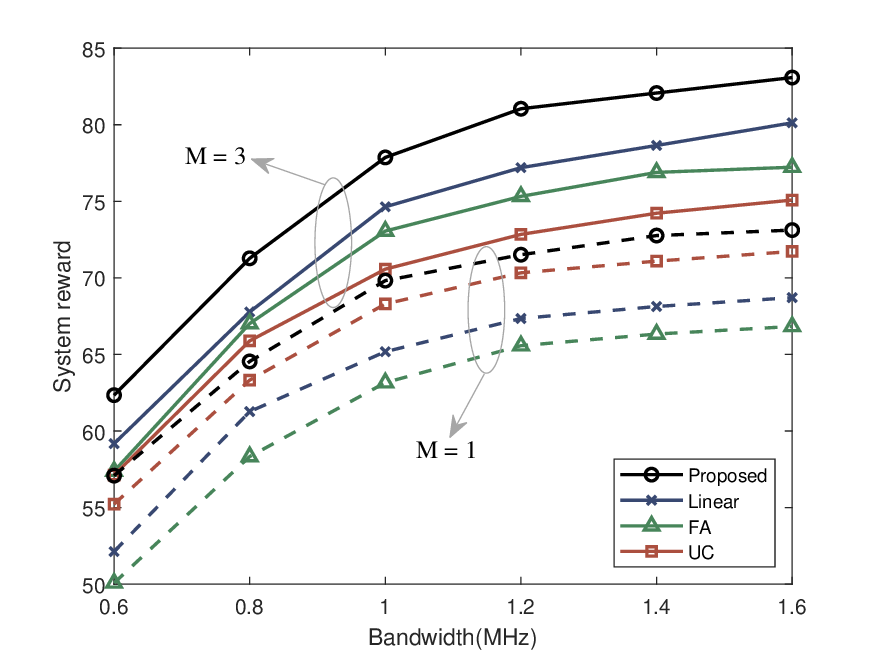}
  \caption{{System reward with varying bandwidth of subcarriers under 3 exit points and 1 exit point.}}
  \label{fig_5}
\end{figure}
The trend of system reward varying with bandwidth of subcarriers $B$ under different number of exit points is represented in Fig. \ref{fig_5}.
Note that only the curves of comparison algorithms whose system reward is relatively close to each other are shown here in order to display the comparison effect more directly. 
{The performance of the proposed algorithm is much better than that of the FTM algorithm, with an average increase of about $25\%$, which will make other curves be compressed very tighty when shown on the figure.
Besides, we find that the impact of changes in bandwidth and compression ratio is not significant when the transmission mode is given, 
therefore we did not demonstrate that.}
When the exit point number $M = 3$, the proposed algorithm has an average performance improvement of {$4.7\%$, $7.3\%$, and $10.3\%$} respectively compared with the Linear algorithm, FA algorithm, and UC algorithm.
As bandwidth becomes increasingly abundant, the impact of optimizing resource-related variables becomes smaller while the linear algorithm is not significantly affected.
Comparing the curves of $M = 3$ and $M = 1$, the performance improvement of the proposed algorithm is $11.8\%$ on average when number of exit points is different. 
The impact of the number of exit points is further enhanced when bandwidth changes.
The system reward increase slowly as the bandwidth increases and trends of different algorithms are slightly different. 
The trend slows down faster when $M = 1$ because computation force overhead of the task is greater when $M = 1$. 
As the bandwidth increases, the computation capacity are limited faster and the reward brought by the bandwidth will be limited faster. 
In conclusion, the proposed algorithm performs better than the comparison schemes with varying bandwidth of subcarriers and the coupling of communication and computation will affect system reward to a great extent.

\begin{figure}[!t]
  \centering
  \includegraphics[scale = 0.6]{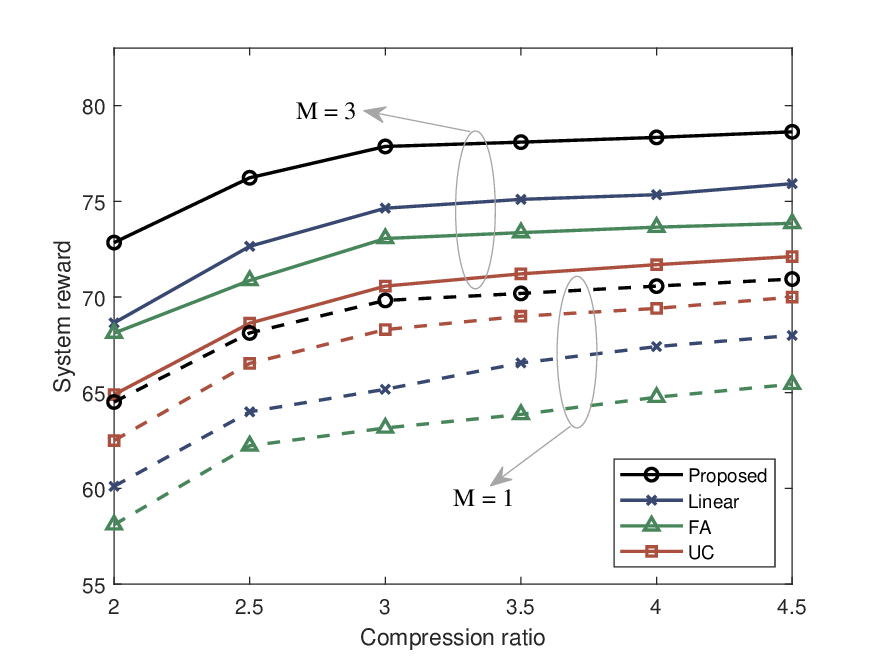}
  \caption{{System reward with varying compression ratio under 3 exit points and 1 exit point.}}
  \label{fig_6}
\end{figure}
We plot the trend of system reward varying with compression ratio under different number of exit points in Fig. \ref{fig_6}.
When the exit point number $M = 3$, the proposed algorithm has an average performance improvement of {$4.5\%$, $6.7\%$, and $10.3\%$} respectively compared with the Linear algorithm, FA algorithm, and UC algorithm.
Comparing the curves of $M = 3$ and $M = 1$, the performance improvement of the proposed algorithm is {$11.6\%$} on average when number of exit point is different. 
The system reward increase is not as significant as when the resource capacity changes and the trend slows down faster. 
The impact of different $M$ on the trend is less significant than the above.
This is partly because the algorithm with a higher compression ratio has better performance under the same accuracy. 
According to the hypothesis of our system, only the impact of changes in the compression ratio itself on delay is considered.
However, an increase in the compression ratio affects the task accuracy and the choice of exit point in practice that also has an impact on system performance. 
It is also partly because when the capacity of various resources is relatively balanced, the compression ratio affects the amount of data transmitted and its change has a smaller impact than the change in resource capacity on resource allocation optimization.
Therefore its influence is weakened on system reward, resulting in a faster slowdown of the curve and a smaller influence of $M$.
Overall, the proposed algorithm performs better than the comparison schemes with varying compression ratio.
The infinite increase in the compression ratio without considering the impact of accuracy does not lead to a corresponding improvement in system model but gradually tends to be constant.

\section{Conclusion}
In this paper, a MEC system for multiple BSs and multiple terminals was proposed, which exploits semantic transmission and {EEoI}.
Based on the semantic transmission process and EEoI mechanism designs, a joint semantic transmission and resource allocation problem was formulated for maximizing delay based system reward.
We decomposed it into three subproblems and designed an efficient BCD based joint semantic transmission and resource allocation algorithm in MEC systems, where 3D matching and convex optimization methods were used to derive optimized solutions.
Simulation results have illustrated that the proposed algorithm significantly improves the delay performance compared with benchmarks. Another interesting finding is that the design of semantic transmission and EEoI during offloading greatly increase system reward, which is more significant compared to other comparisons.
The proposed architecture enables flexibly parameters adjustment and efficient resource utilization, optimizing system reward in intelligent computing scenario.
Future work will focus on the trade-off between the task accuracy and delay in offloading of intelligent computation task, which is a promising direction for further research.

\vfill

\end{document}